\documentclass[a4paper,12pt]{article}
\usepackage{epsfig,amsmath,amssymb}
\usepackage[usenames,dvipsnames]{color}

\newcommand{\beq}{\begin{equation}}
\newcommand{\eeq}{\end{equation}}
\newcommand{\bea}{\begin{eqnarray}}
\newcommand{\eea}{\end{eqnarray}}

\def \p{\partial}
\def \f{\frac}
\def \d{\delta}
\def \L{\mathcal{L}}

\begin{document}

\title{Gauge-invariant matter field actions from iterative N\"other coupling}

\author{Aniket Basu$^{a}$\footnote{Email: aniket.basu@gmail.com}, Parthasarathi Majumdar$^{b}$\footnote{Email: bhpartha@gmail.com} ~and Indrajit Mitra$^{c}$\footnote{Email: indra.theory@gmail.com, imphys@caluniv.ac.in}\\\\
$^{a}$ {\it Department of Physics, Vidyasagar College,}\\{\it 39 Sankar Ghosh Lane, Kolkata 700006, INDIA}\\
$^{b}$ {\it Department of Physics, Ramakrishna Mission Vivekananda University,}\\ {\it Belur Math, Howrah 711202, INDIA}\\
$^{c}$ {\it Department of Physics, University of Calcutta,}\\{\it  92 A.\ P.\ C.\ Road, Kolkata 700009, INDIA.}
}

\date{}

\maketitle

\begin{abstract}
Generalizing Deser's work on pure $SU(2)$ gauge theory, 
we consider scalar, spinor and vector matter fields transforming under arbitrary representations of a non-Abelian, 
compact, semisimple internal Lie group which is a  global symmetry of their actions. These matter fields 
are coupled to Abelian gauge fields through the process of {\it iterative N\"other coupling}. This procedure is 
shown to yield precisely the same locally gauge invariant theory (with the non-Abelian group as the gauge group) 
as obtained by the usual {\it minimal coupling} prescription originating from the Gauge Principle.
Prospects of this non-geometrical formulation,
towards better understanding of physical aspects of gauge theories, are briefly discussed.  
\end{abstract}


\section{Introduction}

To formulate theories of fundamental interactions invariant under local, non-Abelian gauge invariance, the standard 
practice is to appeal to the Gauge Principle (\cite{YM}-\cite{glashow}) 
which is inspired in its turn by the principle of General Covariance 
underlying Einstein's general relativity. According to this principle, physical quantities 
must be gauge invariant, i.e., 
invariant under local gauge transformations. Implementation of the principle leads one to the {\it minimal 
coupling} prescription, under which any partial derivative of a field transforming non-trivially under the action of the 
gauge group, must be augmented by a {\it connection} term. This term compensates for the difference
in gauge transformation property of the field at two different (neighbouring) points.
In general relativity, this prescription is understood in the following 
way \cite{penrose}: Na\"ive parallel transport on a curved surface (which is embedded in a higher dimensional flat
space) of a tangent vector from its initial location, does not
yield a vector  which is tangent to the surface at the new location. We need to make a projection,
of the na\"ively parallel-transported vector, to the tangent space at the new location. 
This is effected by the connection term added on to the partial derivative of the vector.
In gauge theory, the connection term is specified uniquely by the gauge transformation properties of the fields.
Once the augmented (or `covariant') derivatives are constructed and 
{\it curvatures} or field strengths of the gauge field 
are obtained through the Ricci identity, gauge-invariant actions for all fields can be written down.
The Gauge Principle is, thus, a very {\it geometrical} principle.

Starting from the mid-1940s however, many physicists have sought more physical alternatives to this geometrical 
principle (\cite{kraichnan}-\cite{Deser:2017}). Physicists have also questioned whether the Gauge Principle is truly a physical principle, since all dynamical variables in the theory must of necessity be gauge invariant. 
It is thus not clear precisely what new physical information is obtained from the gauge principle, apart from a statement of redundancy of some of the field degrees of freedom used to construct the theory \cite{zee}. 

Further, while the standard formulation has yielded a plethora of physical results all consistent with experimental data 
\cite{peskin}, certain very special physical aspects of non-Abelian gauge interactions, like the anti-screeining
property and asymptotic freedom, 
can only be understood after detailed calculation of 
renormalization-group beta function. 
{\it We aim to understand the more 
complicated physics of non-Abelian gauge theory as a result of simpler constituent dynamics 
based essentially on Abelian gauge invariance and non-Abelian global symmetries.} 
This is the aim of a programme initiated with the present paper.

The alternative approach that we are most motivated by has been proposed by Deser \cite{Deser:1969} (some more field theoretic
works based on \cite{Deser:1969} are \cite{deser:SUGRA}-\cite{0103245}).
The starting point in Deser's work is a Lagrangian with three copies of free Abelian gauge field, 
with the Lagrangian also possessing a global 
$SU(2)$ invariance.  The global symmetry gives rise to a N\"other current for each species of the Abelian gauge field. These 
currents are then coupled to the Abelian gauge fields to generate an additional term in the Lagrangian which again 
is invariant 
under the global symmetry. From this new Lagrangian, one  again constructs N\"other currents, and iterates 
this process until 
such currents cease to be generated. At the point of termination, one ends up with a Lagrangian 
with full $SU(2)$ {\it local} gauge 
invariance. 

In this paper, we show that the procedure discussed above can be generalized to include {\it arbitrary matter fields}: 
starting with a globally $U(1)$ invariant action for
charged scalar, spinor and vector fields, and then proceeding with the iterative N\"other coupling, yields  the same
{\it locally} $U(1)$ gauge invariant matter action as obtained from the minimal coupling prescription pertaining to
local $U(1)$ gauge invariance. This procedure is then generalized to matter actions with global $SU(N)$ symmetry,
with arbitrary representations of the matter fields under $SU(N)$, yielding at the end a locally $SU(N)$
gauge invariant theory for the corresponding matter field.
The number of iterations is always the same as the number of spacetime derivatives needed
to describe the globally symmetric theory.
The procedure thus amply illustrates that {\it the minimal coupling prescription need not be invoked ab initio to 
construct non-Abelian gauge-invariant matter field actions; iterative N\"other coupling achieves the same result.}


Various field theoretic actions have been derived in Refs.\ \cite{deser:SUGRA}, \cite{aragone}, \cite{9605080} and
\cite{0103245} by iterative N\"other coupling (also referred to as self-interaction). But, to our knowledge,
{\it this procedure has not been applied before to derive $U(1)$ and $SU(N)$ gauge-invariant actions for matter fields
of various spins}. 

{\it The motivation for our work is as follows.}
On the one hand, there is a  non-Abelian gauge theory. On the other hand, there is a theory constructed
by starting with Abelian gauge fields and a global non-Abelian symmetry, and then iteratively coupling
the  N\"other current of that symmetry with the Abelian gauge fields.
When the iteration stops, one needs to do an {\it identification} of fields between the two theories to complete the
equivalence. In this way, any non-Abelian gauge theory (with matter fields in our paper) has
been shown here to be classically equivalent to a theory {\it without any explicit non-Abelian
gauge invariance (redundancy)}, but having instead a non-Abelian physical (global) symmetry, and
including all types of realistic matter. This way of interpreting Deser's original program
leads to the remarkable possibility that one can deal with the dynamics of non-Abelian
gauge theories avoiding the mathematical complexity inherent in such theories.

The paper is organized as follows.
In Sec.\ \ref{U(1)}, we obtain the $U(1)$ gauge-invariant Lagrangians involving matter fields with spin 0, spin 1/2 and
spin 1.
In Sec.\ \ref{pg}, we obtain the pure gauge Lagrangian for the $SU(N)$ gauge group, extending Deser's $SU(2)$ gauge group
calculation. Sec.\ \ref{SU(N)} contains our main results leading to the generalization of the iterative N\"other coupling 
procedure to the case of matter fields of spin 0, 1/2 and 1, transforming under arbitrary representations of $SU(N)$ as 
the global 
symmetry group, and demonstrating that the resultant action is identical to the one obtained from `gauging' the appropriate 
matter actions through the minimal coupling prescription. In Sec.\ \ref{concl}, we present our conclusions and outlook.
\section{$U(1)$ gauge invariance for matter field}\label{U(1)}
\subsection{Scalar field}
We start with the Lagrangian of a free complex scalar field and a free Abelian gauge field
\beq
\L_0=(\p_\mu\phi)^*(\p^\mu\phi)-\f{1}{4}F_{\mu\nu}F^{\mu\nu}                   \label{L_0}
\eeq
which is invariant under a global $U(1)$ transformation $\phi\rightarrow\phi e^{ie\omega}$
(in addition to invariance under $U(1)$ gauge transformation of $A_\mu$).
For an infinitesimal transformation,
$\d\phi  = ie\omega\phi$ and 
$\d\phi^*=-ie\omega\phi^*$.
To construct the N\"other current $j^{1\mu}$ (where the superscript 1 stands for the first iteration), we use
\bea
\f{\p\L_0}{\p(\p_\mu\phi)}\d\phi+\f{\p\L_0}{\p(\p_\mu\phi^*)}\d\phi^* = \omega j^{1\mu}    \label{eq1}
\eea
which gives
\bea
j^{1\mu}=ie(\phi\p^\mu\phi^*-\phi^*\p^\mu\phi).
\eea
We add to the Lagrangian the new term 
\bea
\L_1=j^{1\mu} A_\mu.                   \label{L1}
\eea
Then from $\L_1$ we get a further contribution to
the N\"other current, making the replacements
$\L_0\rightarrow \L_1$ and $j^{1\mu}\rightarrow j^{2\mu}$ in (\ref{eq1}):
\bea
j^{2\mu}=2e^2\phi^*\phi A^\mu.
\eea
So we further add
\bea
\L_2=\f{1}{2}j^{2\mu}A_\mu                                     \label{L2}
\eea
to the Lagrangian. Then $\L_2=e^2\phi^*\phi A^\mu A_\mu$, and we see that the factor of $\f{1}{2}$ in (\ref{L2})
is needed to ensure that
\bea
\f{\d}{\d A_\mu}\int d^4x\,\L_2=j^{2\mu}.                    
\eea
As $\L_2$ contains does not contain any derivative of $\phi$, no further contribution to
the N\"other current is generated, and the iteration stops here. Thus the final Lagrangian is
\beq
\L=\L_0+\L_1+\L_2.                                                     \label{Lfinal}
\eeq

It can be easily checked that this final Lagrangian equals (\ref{L_0}) with $\p_\mu$ replaced by 
\bea
D_\mu=\p_\mu+ieA_\mu               \label{Dmu}
\eea
in the scalar part. Thus iterative N\"other coupling has converted the Lagrangian in (\ref{L_0}), which
had only global $U(1)$ invariance in the matter part, to the Lagrangian in (\ref{Lfinal}), in which the matter part
has $U(1)$ gauge invariance.

We also note that $\L_0$, $\L_1$ and $\L_2$ split up the Lagrangian with full $U(1)$
invariance into the propagator, the vertex of order $e$ and the vertex of order $e^2$ respectively.
This will happen in the other cases also.
\subsection{Spinor field}
\beq
\L_0=i\bar\psi\gamma^\mu\p_\mu\psi-\f{1}{4}F_{\mu\nu}F^{\mu\nu}                             \label{Lspinor}
\eeq
is invariant under a global $U(1)$ rotation $\d\psi=ie\omega\psi$,
which gives the N\"other current
\beq
j^{1\mu}=-e\bar\psi\gamma^\mu\psi.
\eeq
We add $\L_1$ as in (\ref{L1}) to $\L_0$. Since $\L_0$ has only a single spacetime derivative of the field,
$j^{1\mu}$ has no field derivative and so, unlike in the previous case, $\L_1$ gives no further N\"other current.
Our final Lagrangian is therefore $\L=\L_0+\L_1$. This equals (\ref{Lspinor}) with $\p_\mu$ replaced by $D_\mu$,
as given by (\ref{Dmu}), in the spinor part.


\subsection{Vector field}
\beq
\L_0=-\f{1}{2}(\p_\mu W_\nu-\p_\nu W_\mu)^*(\p^\mu W^\nu-\p^\nu W^\mu)-\f{1}{4}F_{\mu\nu}F^{\mu\nu} \label{Lvector}
\eeq
is invariant under 
$\d W_\mu=ie\omega W_\mu$ and
$\d W^*_\mu=-ie\omega W^*_\mu$. This gives
\beq
j^{1\mu}=ie\left(W^{\nu *}(\p_\mu W_\nu - \p_\nu W_\mu)-W^\nu(\p_\mu W^*_\nu - \p_\nu W^*_\mu)\right).
\eeq
Then (\ref{L1}) gives
\beq
j^{2\mu}=-e^2\left(2A^\mu W^{\nu *}W_\nu-A^\nu(W_\mu^* W_\nu+W_\nu^* W_\mu)\right).
\eeq
Again (\ref{L2}) and (\ref{Lfinal}) lead us to (\ref{Lvector}) with $\p_\mu$ replaced by $D_\mu$ in the 
part involving the vector field $W_\mu$ \cite{arno}.


\section{$SU(N)$ gauge invariance for pure gauge field}\label{pg}
The calculation of this Section extends Deser's \cite{Deser:1969} calculation, which was done for the
$SU(2)$ group, to the case of the $SU(N)$ group. This is also intended to set the stage for the inclusion
of matter fields.
We start with
\beq
\L_0=-\f{1}{4}(\p_\mu A^a_\nu-\p_\nu A^a_\mu)(\p^\mu A^{a\nu}-\p^\nu A^{a\mu})                      \label{Lgauge}
\eeq
where each $A^a_\mu$ ($a$ running from 1 to $N^2-1$) is an Abelian gauge field. 
This Lagrangian has $U(1)$ gauge invariance $A^a_\mu\rightarrow A^a_\mu+\p_\mu\omega^a$ for each species
of $A^a_\mu$. 
It is also invariant under the global $SU(N)$ transformation
\beq
\d A_\mu^a=gf^{abc}A_\mu^b \alpha^c.                                                                \label{trans}
\eeq
To see this, note that (upto a constant factor) $\L_0$ in (\ref{Lgauge}) equals
${\rm Tr}[(\p_\mu A_\nu-\p_\nu A_\mu)(\p^\mu A^{\nu}-\p^\nu A^{\mu})]$ where the matrix $A_\mu=A^a_\mu T^a$.
(We use the result that ${\rm Tr}[T^a T^b]$ is proportional to $\d^{ab}$,
$T^a$ being the generators of $SU(N)$). So $\L_0$ is invariant under $A_\mu\rightarrow U A_\mu U^\dagger$
where $U$ is a constant $SU(N)$ matrix. Eq.\ (\ref{trans}) is the infinitesimal version of this transformation.

The N\"other current in the first iteration satisfies the relation
\beq
\f{\p\L_0}{\p(\p_\mu A^a_\nu)}\d A_\nu^a=j^{1c\mu}\omega^c\,.
\eeq
This gives us
\beq
j^{1c\mu }=-gf^{abc}(\p^\mu A^{a\nu }-\p^\nu A^{a\mu})A^b_\nu\,.
\eeq
We set up the next term in the Lagrangian as $\L_1=\f{1}{2}j^{1c\mu} A^c_\mu$.
Now (upto a constant factor), $\L_1$ equals ${\rm Tr}[(\p^\mu A^{\nu}-\p^\nu A^{\mu})A_\mu A_\nu]$.
(This can be shown from the results that ${\rm Tr}[T^a[T^b,T^c]]$ is proportional to $f^{abc}$ and that $f^{abc}$
is completely antisymmetric.) So $\L_1$ also is invariant under $A_\mu\rightarrow U A_\mu U^\dagger$.
Therefore we iterate the process once more to obtain
\beq
j^{2e\mu}=g^2f^{abc}f^{ade} A^{b\mu}A^{c\nu} A{^d_\nu}.
\eeq
Then we add
\beq
\L_2=\f{1}{4}j^{2e\mu} A{^e_\mu}
\eeq
so that we satisfy
\bea
\f{\d}{\d A^a_\mu}\int d^4x\,\L_2=j^{2a\mu}.                                \label{factor}
\eea
$\L_2$ is also global $SU(N)$ invariant, as it equals (upto a constant factor) 
${\rm Tr}[[A^\mu,A^\nu][A_\mu,A_\nu]]$.
But as $\L_2$ does not involve derivatives, 
the iteration stops, and the final Lagrangian $\L=\L_0+\L_1+\L_2$ is the familiar $SU(N)$ gauge invariant
Lagrangian
\bea
\L= -\f{1}{4}F{^a_{\mu\nu}}F^{a\mu\nu},
~~F{^a_{\mu\nu}}=\p_\mu A^a_\nu-\p_\nu A^a_\mu-gf^{abc}A{^b_\mu}A{^c_\nu}\,.           \label{LLL}
\eea
It can also be written as $\L=-\f{1}{4} {\rm Tr}[(\p^\mu A^{\nu}-\p^\nu A^{\mu}+ig[A_\mu,A_\nu])^2]$.
The terms of order $g$ and order $g^2$ are respectively $\L_1$ and $\L_2$ deduced above. 

\section{$SU(N)$ gauge invariance for matter fields}\label{SU(N)}
\subsection{Scalar field}
We start with the action
\beq
\L_0=(\p_\mu\phi_i)^*(\p^\mu\phi_i)-\f{1}{4}(\p_\mu A^a_\nu-\p_\nu A^a_\mu)^2        \label{LSUN}
\eeq
where the summation over $i$ is from 1 to the dimension of any representation of $SU(N)$ under which we
want the complex scalar field to transform.
This is invariant under the global $SU(N)$ transformations
\bea
\phi^i\rightarrow \exp\left(ig\alpha^a T^a_{ij}\phi_j\right)\,.
\eea
Using
$\d\phi^i=ig\alpha^a T^a_{ij}\phi_j$ and
$\d\phi^{i*}=-ig\alpha^a \bar{T}^a_{ij}\phi_j^*$
(where the bar over $T$ denotes complex conjugation), we obtain the N\"other current
\beq
j^{1a\mu}=ig\left(T^a_{ij}(\p^\mu\phi_{i}^*) \phi_j-\bar{T}^a_{ij}(\p^\mu\phi_i) \phi_{j}^*\right)\,.
\eeq
We define the next term in the Lagrangian as $\L_1=j^{1a\mu} A^a_\mu$.
In terms of the matrix $A^\mu$ and the column vector $\Phi$ constructed out of $\phi_i$,
$\L_1=ig((\p^\mu\Phi^\dagger)A_\mu\Phi-\Phi^\dagger A_\mu (\p^\mu\Phi))$, which is invariant under
the global $SU(N)$ transformations
$\Phi\rightarrow U\Phi$ and  $A_\mu\rightarrow U A_\mu U^\dagger$. Therefore we generate the next contribution to
the N\"other current
\beq
j^{2b\mu}=g^2(\bar{T}^a_{ij}T^b_{ik}\phi_{j}^*\phi_k+T^a_{ij}\bar{T}^b_{ik}\phi_j\phi_{k}^*)A^{a\mu}.\label{32}
\eeq
This current is of the form $j^{2b\mu}=S^{ab}A^{a\mu}$
where $S^{ab}=S^{ba}$.
Therefore (\ref{factor}) is satisfied when we set 
the next term in the Lagrangian as
\beq
\L_2=\f{1}{2}j^{2b\mu}A^b_\mu.                                                   
\eeq 
This, like $\L_0$ and $\L_1$, is global $SU(N)$ invariant as it 
equals $g^2\Phi^\dagger A^\mu A_\mu\Phi$.
But the iteration stops, and it can be checked that in the final Lagrangian, the matter part of 
the starting Lagrangian (\ref{LSUN}) has been modified into $(D_\mu\phi)_i^*(D^\mu\phi)_i$
where
\bea
(D_\mu)_{ij}=\d_{ij}\p_\mu+igA^a_\mu(T^a)_{ij},                                       \label{DDmu}
\eea
so that we have arrived at $SU(N)$ gauge invariance.
Writing in the matrix form $(D_\mu\Phi)^\dagger(D^\mu\Phi)$, where $D_\mu=\p_\mu+igA_\mu$,
we find that the terms of order $g$ and order $g^2$ are respectively $\L_1$ and $\L_2$ deduced above.

An important point is that the pure gauge part of $\L_0$ in (\ref{LSUN}) also generates N\"other current due to invariance
under (\ref{trans}). But since the corrections $\L_1$ and $\L_2$ to the matter field Lagrangian
do not involve derivatives of $A_\mu^a$, the iterative N\"other coupling from the pure gauge part
proceeds simultaneously with (as the same parameters $\alpha^a$ are involved), but independent of,
the iterative N\"other coupling from the matter part. 
So together with $SU(N)$ gauge invariance in the matter part, we end up with (\ref{LLL}) as in Section \ref{pg}.
This will happen for the spinor and the vector fields also.

For completeness, we note that the calculations of this section are easily modified when the matter field
transforms under the adjoint representation of $SU(N)$. 
As this representation is real, we start with $N^2-1$ species of real scalar field $\phi^a$. 
The global $SU(N)$ invariant Lagrangian is
\beq
\L_0=\frac{1}{2}(\p_\mu\phi^a)(\p^\mu\phi^a)-\f{1}{4}(\p_\mu A^a_\nu-\p_\nu A^a_\mu)^2\,.
\eeq
The generators have the elements
$(T^b)_{ac}=if_{abc}$
and so the scalar field transforms as $\d\phi^a=g f^{abc}\phi^b\alpha^c$.
The currents from $\L_0$ and $\L_1=j^{1a\mu} A^a_\mu$ are
\bea
j^{1c\mu}=gf^{abc}(\p^\mu\phi^a)\phi^b\,,\\
j^{2e\mu}=g^2f^{abc}f^{ade}\phi^bA^{c\mu}\phi^d\,.
\eea
Then adding $\L_2=\f{1}{2}j^{2e\mu}A^e_\mu$ gives the $SU(N)$ gauge invariant Lagrangian which
contains the covariant derivative $(D_\mu)_{ac}=\d_{ac}\p_\mu+igA_\mu^b(T^b)_{ac}$ in the matter part.

\subsection{Spinor field}
\beq
\L_0=i\bar\psi_i\gamma^\mu\p_\mu\psi_i -\f{1}{4}(\p_\mu A^a_\nu-\p_\nu A^a_\mu)^2
\eeq
is invariant under the global $SU(N)$ rotation
$d\psi_i=ig\alpha^aT^a_{ij}\psi_j$, giving $j^{1a\mu}=-g\bar\psi_i\gamma^\mu T^a_{ij}\psi_j$
and $\L=\L_0+j^{1a\mu}A^{a}_\mu$. This modifies the matter part in $\L_0$ with $\p_\mu$ replaced by
$D_\mu$ as in (\ref{DDmu}).






\subsection{Vector field}
\beq
\L_0=-\f{1}{2}(\p_\mu W_{i\nu}-\p_\nu W_{i\mu})^*(\p^\mu W^{\nu}_i-\p^\nu W^{\mu}_i)
     -\f{1}{4}(\p_\mu A^a_\nu-\p_\nu A^a_\mu)^2 
\eeq
is invariant under
$\d W_i^\mu=ig\alpha^a T_{ij}^a W_j^\mu$ and
$\d W_i^{\mu *}=-ig\alpha^a T_{ij}^a W_j^{\mu *}$.
This gives the current
\beq
j^{1a\mu}=ig
\left(\bar{T}^a_{ij} W_{j\nu}^{*}\left(\p^\mu W^{\nu}_i-\p^\nu W^{\mu}_i\right) -
T^a_{ij} W_{j\nu}\left(\p^\mu W^{\nu *}_i-\p^\nu W^{\mu *}_i\right)\right)\,.
\eeq
Then setting $\L_1=j^{1a\mu} A^a_\mu$ gives
\bea
j^{2b\mu}&=&-g^2 \left(\bar{T}^a_{ij} T^b_{ik}W_j^{\nu *} W_{k\nu}+T^a_{ij}\bar{T}^b_{ik} W_{j}^\nu W_{k\nu}^*\right)A^{a\mu}
\nonumber    \\
&&+g^2 \left(\bar{T}^a_{ij} T^b_{ik}W^{\mu *}_j W_{k\nu}+T^a_{ij}\bar{T}^b_{ik}W{_j^\mu} W_{k\nu}^*\right)A^{a\nu}\,.
\label{43}
\eea
This current is of the form $j^{2b\mu}=S^{ab}A^{a\mu}+{S'^{ab\mu}}_\nu A^{a\nu}$
where $S^{ab}=S^{ba}$ and $S'^{ab\mu\nu}=S'^{ba\nu\mu}$.
This ensures that (\ref{factor}) is satisfied when we set $\L_2=\f{1}{2}j^{2b\mu} A^b_\mu$. 
Again one can check that $\L_0+\L_1+\L_2$ has  the covariant derivative (\ref{DDmu})
in the matter part of the Lagrangian.

\section{Conclusions and outlook}\label{concl}
Our paper is more than an explicit completion of the Deser program for non-Abelian 
gauge theories to include matter field sources.
We have an alternative motivation for this
apparent completion, as follows.
The foregoing sections establish that non-Abelian gauge invariance of very general classes of field theories is a 
{\it dynamical consequence} of the iterative N\"other coupling procedure, where one has put in  only the 
non-Abelian global symmetries and Abelian gauge invariance.
If one is able to derive all dynamical 
results -- classical and quantum -- of non-Abelian gauge theories based on these invariances alone,
one may not need to
consider the full non-Abelian gauge invariance with all its complications.
In that case,
the property of asymptotic freedom (attributed to non-Abelian gauge field self-interactions)
may be traced to a somewhat different physical origin. It can then become clearer why it is only the self-interactions 
of non-Abelian gauge fields that possess this property,
in contrast to the entire gamut of fundamental interactions which are not asymptotically free.
Also, within such a formulation,
{\it infrared} strong-coupling phenomena like the phase transition to a quark-gluon plasma, 
quark confinement, and low energy
hadron physics in general, might become easier to handle.
One can even hope for a scenario in which Abelian gauge fields are used on a lattice 
for extracting non-perturbative dynamical information like hadron masses and decay widths.

Another motivation for this programme arises from the fact that Abelian gauge theories have already been 
formulated in terms of 
gauge-{\it invariant} fields, within the so-called `gauge-free' approach \cite{srijit}, 
using the unique and natural projection
operator given in the $U(1)$ gauge field action itself. 
Our current underpinning in this paper of non-Abelian gauge theories 
on Abelian gauge fields, may afford us a way to do this for non-Abelian gauge fields as well.

\section*{Acknowledgments}
I.M.\ thanks the UGC (DRS) program of the Department of Physics, University of Calcutta, for support. 
P.M.\ thanks Romesh Kaul for useful discussions. The authors also thank an
anonymous referee for bringing the paper cited in \cite{arno} to our attention.

\end{document}